\documentclass[sigconf,screen]{acmart}
 
\AtBeginDocument{%
  }

\copyrightyear{2026}
\acmYear{2026}
\setcopyright{cc}
\setcctype{by}
\acmConference[FAccT '26]{The 2026 ACM Conference on Fairness, Accountability, and Transparency}{June 25--28, 2026}{Montreal, QC, Canada}
\acmBooktitle{The 2026 ACM Conference on Fairness, Accountability, and Transparency (FAccT '26), June 25--28, 2026, Montreal, QC, Canada}
\acmDOI{10.1145/3805689.3812355}
\acmISBN{979-8-4007-2596-8/2026/06}

\usepackage{listings}
\usepackage{subfig}

\begin{document}

\title{The DSA's Blind Spot: Algorithmic Audit of Advertising and Minor Profiling on TikTok}

\author{Sara Solarova}
\affiliation{%
  \institution{Kempelen Institute of Intelligent Technologies}
  \city{Bratislava}
  \country{Slovakia}
}
\email{sara.solarova@kinit.sk}
\orcid{0000-0003-2800-3572}

\author{Matej Mosnar}
\affiliation{%
  \institution{Kempelen Institute of Intelligent Technologies}
  \city{Bratislava}
  \country{Slovakia}}
\email{matej.mosnar@kinit.sk}
\orcid{0009-0009-2466-7225}

\author{Matus Tibensky}
\affiliation{%
  \institution{Kempelen Institute of Intelligent Technologies}
  \city{Bratislava}
  \country{Slovakia}}
\email{matus.tibensky@kinit.sk}
\orcid{0000-0001-9928-3474}

\author{Jan Jakubcik}
\affiliation{%
  \institution{Kempelen Institute of Intelligent Technologies}
  \city{Bratislava}
  \country{Slovakia}}
\email{jan.jakubcik@kinit.sk}
\orcid{0009-0009-5162-2858}

\author{Adrian Bindas}
\affiliation{%
  \institution{Kempelen Institute of Intelligent Technologies}
  \city{Bratislava}
  \country{Slovakia}}
\email{adrian.bindas@kinit.sk}
\orcid{0009-0002-8796-6475}

\author{Simon Liska}
\affiliation{%
  \institution{Kempelen Institute of Intelligent Technologies}
  \city{Bratislava}
  \country{Slovakia}}
\email{simon.liska@kinit.sk}
\orcid{0009-0008-0873-9797}

\author{Filip Hossner}
\affiliation{%
  \institution{Kempelen Institute of Intelligent Technologies}
  \city{Bratislava}
  \country{Slovakia}}
\email{filip.hossner@kinit.sk}
\orcid{0009-0005-6630-0158}

\author{Matúš Mesarčík}
\affiliation{
  \institution{Kempelen Institute of Intelligent Technologies}
  \city{Bratislava}
  \country{Slovakia}
}
\affiliation{
  \institution{Comenius University Bratislava}
  \city{Bratislava}
  \country{Slovakia}
}
\email{matus.mesarcik@kinit.sk}
\orcid{0000-0003-3311-5333}

\author{Ivan Srba}
\affiliation{%
  \institution{Kempelen Institute of Intelligent Technologies}
  \city{Bratislava}
  \country{Slovakia}
}
\email{ivan.srba@kinit.sk}
\orcid{0000-0003-3511-5337}

\renewcommand{\shortauthors}{Solarova et al.}

\begin{abstract}
Adolescents spend an increasing amount of their time in digital environments where their still-developing cognitive capacities leave them unable to recognize or resist commercial persuasion. Article 28(2) of the Digital Service Act (DSA) responds to this vulnerability by prohibiting profiling-based advertising to minors.  However, the regulation's narrow definition of ``advertisement'' excludes current advertising practices including influencer paid partnerships and brand promotional content that serve functionally equivalent commercial purposes. We provide the first empirical evidence of how this definitional gap operates in practice through an algorithmic audit of TikTok. Our approach deploys sock-puppet accounts simulating a pair of minor and adult users with matching interest profiles. The content recommended to these users is automatically annotated, enabling systematic statistical analysis across four video categories: containing formal, disclosed, undisclosed advertisement and non-advertisement; as well as advertisement topical relevance to user's interest. Our findings reveal a stark regulatory paradox. TikTok demonstrates formal compliance with Article 28(2) by shielding minors from profiled formal advertisements, yet both disclosed and undisclosed ads exhibit significant profiling aligned with user interests (5-8 times stronger than for adult formal advertising). The strongest profiling emerges within undisclosed commercial content, where creators/brands fail to label paid partnership/promotional content and the platform neither corrects this omission nor prevents its personalized delivery to minors. These results demonstrate that minors remain exposed to algorithmically targeted commercial content through the same recommendation mechanisms the DSA seeks to constrain. We argue that protecting minors requires expanding the definition of advertisement in EU law to encompass influencer and brand promotional content, and  ensuring that any such expansion is accompanied by a corresponding prohibition on profiling-based targeting of minors, so that commercial content cannot circumvent protections merely by operating outside formal advertising channels.
\end{abstract}

\ccsdesc[500]{Human-centered computing~HCI design and evaluation methods}
\ccsdesc[300]{Human-centered computing~Collaborative and social computing design and evaluation methods}
\ccsdesc[500]{Social and professional topics~Computing / technology policy}

\keywords{Digital Services Act, advertisement, algorithmic auditing, minor profiling, TikTok}


\maketitle

\section{Introduction}

A teenager on TikTok today interacts in a fundamentally different information environment than most adults did when they first opened social media. A decade ago, feeds had natural limits. Users saw what their contacts posted, and when posting stopped, so did the feed. Today, content loads immediately as an endless stream, shaped by everything a user has ever watched, liked or lingered on. There is effectively no moment when the feed runs out, no natural stopping point. Nor do users have meaningful visibility into what the algorithm has learned about them and how it operates. And within this endless stream, the line between entertainment and advertisement (hereafter also ads) has effectively disappeared.

For adults, this blurring is confusing. For children, it may be not only problematic but also harmful. Decades of research, on food marketing to children \cite{harris2009food, cairns2013systematic, boyland2015food}, gambling-like design of the apps and adolescent behaviour \cite{rixen2023loop, montag2019addictive}, adolescent brain development \cite{steinberg2004risk, crone2018media, maza2023association} and minors advertising literacy \cite{rozendaal2011reconsidering, van2017advertising, van2017processes, packer2022advertising, john1999consumer, nairn2008s} demonstrate that minors lack the cognitive defences against ads that adults have developed. 

Platforms like TikTok are where these vulnerabilities meet commercial interests. TikTok has 169 million monthly active users in the EU \cite{tiktok_dsa_transparency_2025}, and among young Europeans it has become a primary channel for information: 64 percent of 15-24 year-olds now cite social media as their main source of information overall \cite{eurobarometer2025socialmedia}. In 2024, TikTok generated approximately €21 billion in revenue and the vast majority was from advertising \cite{businessofapps2024tiktok}. Attention has become the commodity, where advertising is the revenue and the algorithmic feed exists to maximise both. The algorithm infers interests, insecurities, and aspirations from behavioural signals invisible to anyone but the platform. Minors no longer scroll through content from friends but from brands and influencers, much of it packed with ads, sometimes disclosed but often undisclosed. While the platforms provide disclosure tools, creators rarely use them, often burying commercial indicators in hashtags effectively invisible to users. At the EU level, only 20 percent of influencers systematically indicate the commercial nature of their content \cite{europeancommission2024influencer}.

The European Union has recognised these harms in the Digital Services Act (DSA), which entered into force in 2024 and introduced Article 28(2), which prohibits online platforms, including Very Large Online Platforms (VLOPs) and Very Large Search Engines (VLOSEs), from presenting ads based on profiling when they know ``with reasonable certainty'' that the user is a minor. The intent is clear: minors shall not be targeted with personalised advertising. But the prohibition contains a critical gap. Article 3(r) of the DSA defines ``advertisement'' narrowly, covering only ads purchased through a platform's ad system for which a platform is directly paid (hereafter: formal ads) \cite{EU_DSA_2022_2065}. This definition does not capture brand or influencer marketing, affiliate content, or unlabelled commercial posts, which are precisely the formats that dominate what minors actually see.

Minors should be able to participate in online spaces without being manipulated, exploited, or exposed to risks they cannot  defend themselves against the way adults can. The DSA's Article 28(2) embodies this principle but principles without enforcement mechanisms are merely aspirational. This raises two questions: first, how to verify that platforms are honouring their obligations and second, whether the regulatory framework covers the commercial practices that actually reach children.

Answering these questions presents a fundamental methodological challenge. Algorithmic systems are not static objects that can be inspected once and declared compliant based on a document review or management attestations \cite{SolarovaEtAl2026DSA}. They are dynamic, adaptive, and opaque, commonly understood as black boxes whose behaviour can only be assessed empirically. We must study these systems behaviourally, longitudinally, through systematic simulation and observation of their outputs that reveal what the algorithm actually does rather than what the platform claims it does. However, the research on verifying compliance with provisions like Article 28(2) is limited or completely absent.

To fill this gap, we conduct an \textit{algorithmic audit} on TikTok using sock-puppet accounts simulating pairs of minor and adult users with the same interest profiles. The observed recommended videos are then, with the help of Large Vision-Language Models (LVLMs), automatically categorised as containing:
\begin{enumerate}
    \item \textit{Formal ads} -- commercial content promoted through TikTok's advertising system, where a creator pays the platform for distribution, this satisfies Article 3(r)'s narrow definition of advertisement as content promoted ``against remuneration specifically for promoting that information'';
    \item \textit{Disclosed ads} -- commercial content where remuneration does not flow to the platform, labelled by creators using TikTok's disclosure tools. This includes \textit{paid partnership} (where a brand pays a creator to promote products on the creator's account) and \textit{promotional content} (where a brand promotes its own products on its own account or where an individual promotes a brand they own or have a commercial stake in, without paying the platform or any third party);
    \item \textit{Undisclosed ads} -- same type of content as commercial content—paid partnerships or brand self-promotion, where creators have failed to apply disclosure labels despite platform rules requiring it; and finally, 
    \item \textit{Non-ads} -- content with no commercial purpose.
\end{enumerate}

Furthermore, the topic of advertisement is detected and compared with the user's interest -- such a match serves as an input to measuring the potential personalization and profiling. We assess whether the recommendation algorithm treats these advertisement types differently and specifically, whether profiling occurs for commercial content that falls outside the DSA's narrow definition of ``advertisement''.

We focus on TikTok for the following reasons. First, it is the platform where the gap between regulatory intent and commercial reality is most visible: when a user opens TikTok, they land by default on the ``For You Page'', which is a feed driven entirely by the platform's recommender system rather than social connections. Most users spend the majority of their time on this algorithmically curated feed, meaning that content minors see has been selected based on their inferred preferences and interests. Second, TikTok has been the subject of multiple EU enforcement actions under the Digital Services Act, including proceedings concerning addictive design and default settings for minors \cite{EC2025TikTokAdTransparency}, making it a critical test case for whether existing regulation can protect minors from commercial exploitation \cite{bik2024report}.

Our findings reveal a significant enforcement gap. Formal advertisements are recommended to minors to a limited extent (in comparison with adults) and show no evidence of profiling, consistent with DSA compliance. However, disclosed, and especially undisclosed commercial content (which represents a substantial share of commercial content minors are exposed to), show clear evidence of profiling based on inferred user interests. In other words, while the minors are exposed to less amount of commercial content, such content is highly tailored/personalized according to their profiles. The implication is stark: the DSA protects minors from one category of commercial content while leaving the dominant forms of commercial exposure practically unregulated.

Our main contributions are as follows:
\begin{itemize}
\item We introduce a novel methodology proceeding from the sock-puppet algorithmic audits. It can be characterized as a \textit{paired user} study design (as an analogy to \textit{paired ads} methodology \cite{10.1145/3715275.3732172}), which provides a unique user-centric insight what kind of commercial content is recommended to a minor/an adult under the same conditions (an interest, a time period, a gender, a location) removing potential confounding factors.
\item By applying the proposed methodology to 3 user pairs over the period of 10 days, we collected and automatically evaluated 7095 TikTok videos, of which 1346 were classified as containing one of studied advertisement types. Based on this data, we provide the first empirical evidence that profiling-based targeting of minors continues on TikTok through content formats not covered by DSA Article 3(r)'s definition of ``advertisement'' (including disclosed and undisclosed influencer and branded content).
\item We identify a regulatory gap: while the DSA prohibits profiling-based advertising to minors for formal ads, influencer and brand promotional content falls outside this definition entirely, leaving the dominant forms of commercial exposure minors face without any corresponding prohibition on profiling-based targeting. To address this gap, we propose four specific policy recommendations.
\end{itemize}

\section{Background and Related Work}

\subsection{Negative Influence of Social Media Platforms and Advertisement on Minors}

Adolescence is a critical developmental period during which ongoing brain development increases vulnerability to commercial content and its influence. Regions of the brain governing reward and salience, the ventral striatum, amygdala, and insula, become hyperactive while the prefrontal cortex, which governs cognitive control and self-regulation, continues to mature into the mid-twenties \cite{crone2018media, steinberg2004risk}. The prefrontal cortex enables deliberate evaluation of incoming information, including recognising and resisting persuasive intent, and its immaturity during adolescence means that cognitive control is substantially limited at the moment it is most needed.

This matters because the platform environment described above is one in which commercial and organic content are frequently indistinguishable, particularly where creator-driven commercial content carries no disclosure label. Algorithmically-curated feeds, infinite scrolling, and sentiment-aware recommendation systems create a continuous, non-predictable stream of content that sustains engagement through reward patterns theoretically comparable to partial reinforcement in gambling \cite{rixen2023loop, montag2019addictive}, with no natural stopping point at which a user can step back and assess what they have been watching.

Resisting advertising requires recognising persuasive intent and the capacity to resist it \cite{harris2009food, friestad1994persuasion}. Research shows that while minors can identify advertising as selling products, they often fail to recognise its intent to shape attitudes and behaviour, a limitation that persists through adolescence \cite{packer2022advertising, steinberg2004risk}. Even when recognised, active resistance requires switching from automatic affective processing to deliberate, critical evaluation, a capacity dependent on still-developing prefrontal function \cite{crone2018media}. Because this switching capacity is substantially limited, adolescents process advertising through low-elaboration pathways. Rather than engaging in critical evaluation, they respond to advertising affectively: commercial content that matches their interests evokes positive feelings, and these feelings transfer directly to brand attitudes through processes that bypass deliberate evaluation \cite{van2017processes, rozendaal2011reconsidering}. Advertising literacy alone is therefore insufficient, as its application depends on capacities that are still developing.

Profiling-based targeting exploits this constraint directly. Minors cannot recognize targeting as a deliberate marketing technique and therefore they process such advertising in a non-critical manner \cite{van2017advertising, van2017processes}. As a result,targeting increases advertisement liking while evading recognition and foreclosing the first condition of any defence model before adolescents even encounter the content.

Creator-driven commercial content compounds these vulnerabilities through a mechanism rooted in the same social reward hyperactivity that characterises adolescent brain development. Adolescents' heightened sensitivity to peer-related social signals makes them particularly susceptible to para-social relationships, the illusion of a long-lasting personal bond with a media figure that fosters trust and reduces critical distance \cite{horton1956mass, berryman2017guess}. Influencers function as quasi-peers whose opinions carry persuasive weight precisely because of this dynamic \cite{berryman2017guess}, and social learning theory indicates that liking a figure increases the probability of imitating their actions \cite{coates2019effect}. Although disclosures can activate recognition of advertising intent \cite{boerman2020disclosing}, minors with strong para-social relationships do not translate that recognition into critical evaluation — the bond with the creator outweighs the scepticism that awareness of commercial intent would otherwise trigger \cite{boerman2020disclosing}. When content carries no disclosure at all, sponsored material mimics the visual aesthetics of everyday posts and recognition is foreclosed entirely \cite{gurkaynak2018navigating, musiyiwa2023sponsorship}. The consequences are documented: internal research disclosed by former Facebook employee Frances Haugen confirmed that platforms were aware that exposure to appearance-related content worsens body image and self-esteem among teenage girls, yet algorithmic amplification of such content continued \cite{Wells2021Facebook}. It is within this context of structural vulnerability that the adequacy of the current regulatory framework must be assessed. 

\subsection{Regulation of Displaying Advertisements to Minors in the DSA}
\label{sec:dsa}

DSA establishes a tiered framework of obligations for digital services operating in the EU \cite{Husovec2024PrinciplesDSA}. Universal obligations apply to all intermediary services, while the specific subset of hosting services -- online platforms -- face additional requirements covering, among other things, the protection of minors, recommender systems, and advertising transparency. Very Large Online Platforms (VLOPs) face the strictest requirements, including mandatory risk assessments, mitigation measures, and independent external audits.

The reach of these obligations depends fundamentally on how the DSA defines ``advertisement''.  Article 3(r) DSA defines advertisement as ``information designed to promote the message of a legal or natural person, irrespective of whether to achieve commercial or non-commercial purposes, and presented by an online platform on its online interface against remuneration specifically for promoting that information''. This definition is grounded in payment structure rather than commercial purpose, meaning only content for which a platform receives remuneration for promotion qualifies as an advertisement under the DSA. Commercial content presented without such remuneration, including hidden or undisclosed commercial content, falls outside its scope \cite{Husovec2024PrinciplesDSA}.

This boundary directly determines the scope of minors' protection under the DSA. Article 26(1) DSA requires platforms to ensure that recipients can identify each advertisement as such, including the entity behind it, who paid for it, the main targeting parameters used, and how users can change those parameters. Article 26(2) DSA extends this by requiring platforms to provide creators with a functionality to declare when content constitutes a commercial communication, placing compliance responsibility on creators rather than platforms. Article 26(3) independently prohibits profiling-based targeting using special categories of personal data regardless of the recipient's age. For minors, Article 28(2) DSA goes further, prohibiting platforms from presenting profiling-based advertisements to users they are "aware with reasonable certainty" to be minors, where the European Commission's Guidelines define minors as individuals below 18 years \cite{EuropeanCommission2025MinorsDSA}. Because Articles 26(1) and 28(2) operate within Article 3(r)'s definition, their obligations do not extend to commercial content for which the platform receives no remuneration.

The application of Article 28(2) further depends on the GDPR's definition of profiling. Some authors argue that where data is voluntarily disclosed by the user, its subsequent use falls outside the scope of profiling and therefore outside Article 28(2) \cite{WoltersBorgesius2025}. In our view, this argumentation is incorrect. Profiling under GDPR Article 4(4) is defined as "any form of automated processing of personal data consisting of the use of personal data to evaluate certain personal aspects relating to a natural person, in particular to analyse or predict aspects concerning that natural person's performance at work, economic situation, health, personal preferences, interests, reliability, behaviour, location or movements" \cite{EU_GDPR_2016_679}. This definition emphasises what is done with data rather than how it was acquired. The European Data Protection Board (EDPB) affirms this position by treating profiling as a function of automated logic rather than data origin \cite{WP251rev01_2018}, and its guidance on age-assurance warns controllers against using data collected for age verification to profile minors beyond what is strictly necessary \cite{EDPB2025AgeAssurance}.

The narrow definition of advertisement by Article 3(r) is partially addressed by other EU instruments. The Unfair Commercial Practices Directive (UCPD) covers deceptive and manipulative commercial practices against consumers regardless of payment structure \cite{ClausenRiefa2019}, including hidden marketing (UCPD, Annex I, point 11) and false representations of commercial intent (UCPD, Annex I, point 22). Article 5(3) UCPD requires that disclosures to be understandable to minors as a vulnerable consumer group,  while Annex I point 28 prohibits directly prompting minors to purchase advertised products \cite{EUCommission2021UCPDGuidance}. Digital consent by children is further governed by Article 8 GDPR \cite{MacenaiteKosta2017}, and the Audiovisual Media Services Directive (AVMSD) imposes protection requirements on video-sharing platforms \cite{EuropeanParliament2025MinorsOnline}. None of these instruments, however, specifically imposes on platforms a systemic obligation to detect or prevent the profiling-based delivery of commercial content to minors. The Article 28(2) prohibition therefore remains the most specific and enforceable protection available in this domain, yet its reach is confined to the narrow category of content that Article 3(r) recognises as advertisement.

\subsection{Algorithmic Auditing}

Algorithmic audits are commonly defined as systematic assessments of the negative impacts that algorithmic systems may have on the rights and interests of affected stakeholders, together with identifying of algorithmic features or situations that give rise to such harms~\cite{10.1177/2053951720983865}. In the context of social media platforms, AI-driven recommender systems, including ad delivery algorithms, are largely opaque and inaccessible to external inspection. Audits assessing their properties (e.g., potential biases) must therefore rely on behavioural analysis. Rather than examining internal model parameters, researchers infer system properties from platform responses to controlled user interactions, such as content consumption or engagement, and by analysing the resulting recommendations.

To operationalize such behavioural audits, prior work has primarily employed simulated user interactions through automated agents (bots)~\cite{yaudit-recsys} or human participants~\cite{10.1145/3351095.3372879}, commonly referred to as sock-puppeting audits and crowdsourcing (or collaborative) audits, respectively~\cite{Sandvig2014Audits}. These approaches allow researchers to systematically probe platform behaviour under predefined conditions, such as specific user interests, interaction patterns, or demographic profiles. When deployed at sufficient scale and diversity, algorithmic audits have proven effective in uncovering issues related to harmful content amplification (e.g., misinformation \cite{srba2023auditing}), political bias, or unfair exposure, and they can also serve as an independent mechanism for validating regulatory compliance, including audit reports mandated under the DSA for VLOPs and VLOSEs \cite{SolarovaEtAl2026DSA}.

A typical algorithmic audit follows a structured multi-step process. First, an audit question is formulated, which guides the design of audit scenarios capturing specific user profiles and sequences of actions intended to simulate realistic user behaviour. These scenarios are then executed by bots or human agents, during which platform responses (such as recommended content or changes in content exposure) are systematically recorded. Finally, the collected data are analysed to detect the presence, magnitude, and dynamics of the audited phenomenon, enabling empirical conclusions about the platform’s algorithmic behaviour and its potential societal impacts.

The current generation of algorithmic audits is still in early stages and may suffer from various limitations, such as low reproducibility and short-term validity of results~\cite{10.1145/3726302.3730293,simko2021continuous-automatic-audits}; or focus on few online platforms, Western contexts, particularly the US, English language data and specific (commonly over-simplified) user properties~\cite{10.1145/3715275.3732026}. However, due to growing research efforts as well as new paradigms, such as model-based algorithmic auditing~\cite{srba2025ewaf}, their overall maturity and robustness is continuously increasing.

\subsection{Auditing Policy Compliance and Ad Delivery Algorithms}

To assess platforms' compliance with the DSA, VLOPs and VLOSEs are required to undergo annual external audits. These audits aim to provide an independent assessment of compliance with legal obligations, particularly regarding risk assessment and mitigation of systemic risks (DSA Articles 33, 34, and 37). Analysis of public audit reports showed that such external audits rely mostly on conventional audit methodologies, rooted in compliance auditing traditions developed for financial reporting and IT controls \cite{SolarovaEtAl2026DSA}. These methodologies were, however, designed for relatively static systems and therefore face unprecedented challenges when applied to modern AI-based systems that interact dynamically with human users. 

In parallel, there are complementary independent audits by researchers, non-profit watchdog organizations or algorithmic assessment / ethics-focused companies. These audits commonly apply various algorithmic auditing methodologies. \citet{10.1145/3715275.3732052} recently conducted a literature review of audit study designs suitable for investigating algorithmic-driven risks in digital services focusing on DSA provisions addressing algorithmic systems. The outcome is four main categories of methodologies: risk-uncovering, reverse engineering, interface design, and risk-measuring; with the main steps and best practices associated with each approach.

While existing literature proves that various types of algorithm audits represent a suitable approach how to assess compliance with policy and regulations, the number of existing studies remain limited. In regards to black-box audits of ad delivery algorithms, the existing works (such as~\cite{10.1145/3715275.3732172}) employ \textit{paired ads} methodology. In this methodology, an auditor runs a pair of ads targeting the same audience and at the same time~\cite{10.1145/3715275.3732172}. The content of ads, however, intentionally introduces a hypothetical bias (e.g., one ad is hypothesized that will be propagated more to men and second to women). Algorithmic bias is then measured by looking at relative difference in delivery along a demographic attribute of interest to the auditor. 

Such an \textit{ad-centric} approach (centred around a specific ad analysing how it is delivered to users with various demographics) is, however, not appropriate for assessing the compliance with protection of minors, which is subject of our study. Instead, we aim for more \textit{user-centric} approach, which will allow us to focus on a user (e.g., a minor or an adult) and study what kind of advertisements are delivered (in terms of type and topic) to such a user. To the best of our knowledge, there is no such work yet. To make it possible, the sock-puppeting algorithmic audits represents a potential and suitable solution, which is adopted also in our work.

\section{Preliminary Analysis}

As existing works focusing on user-centric algorithmic auditing of ad delivery systems remain limited or even completely absent as well as findings from the existing studies on constantly-evolving TikTok ad delivery system may be already obsolete (also as a result of DSA adoption), we opted for iterative approach. At first, we applied qualitative methods to obtain initial insights, namely a \textit{documentary analysis} of TikTok's disclosure framework and a \textit{manual content analysis} from selected creators. The obtained initial findings subsequently served us as important inputs to design a quantitative methodology and experiment utilizing the sock-puppeting algorithmic audit (e.g., to determine topics the simulated users should express interest in, or to determine the taxonomy of ad types present on the TikTok platform).

\subsection{Documentary Analysis}
At first, we analysed TikTok's publicly available policies on commercial content disclosure using four sources: the Branded Content Policy \cite{tiktok_branded_content_policy_2025}, the Commercial Content Disclosure guidelines in TikTok's Business Help Center \cite{tiktok_content_disclosure_setting_2025}, the Community Guidelines \cite{tiktok_community_guidelines_2025}, and the TikTok’s Commercial Content Library \cite{tiktok_library_faq_2025}.

The Branded Content policy (BCP) \cite{tiktok_branded_content_policy_2025} defined branded content as content promoting third-party brand in exchange for payment or incentive for the creator, such as gifted products, monetary payment, affiliate commissions and brand partnerships. TikTok shifts compliance responsibility onto creators, who must ensure adherence to applicable laws and platform rules.

When creators post branded content, they must enable the 'Disclose commercial content' toggle. The toggle offers two options: ``Your brand'' or ``Branded content.'' Selecting ``Your brand'' produces a ``Promotional content'' label; selecting ``Branded content'' produces a "Paid partnership" label. Once published, these labels cannot be changed and if there is a mistake creators must delete and repost to correct errors. Advertisements purchased directly through TikTok Ads Manager carry a separate ``Sponsored'' or ``Ad'' label (depending on whether it is TikTok app or desktop version) that TikTok injects automatically \cite{tiktok_branded_content_policy_2025}.

TikTok's Community Guidelines \cite{tiktok_community_guidelines_2025} address commercial disclosure under "Regulated Goods, Services, and Commercial Activities." The guidelines state disclosure is required when ``marketing for your own business, product, or service'' or when engaging in ``promotion or review of a third-party brand, product, or service in exchange for money, goods, or any other kind of incentive''. However, the guidelines remain broad regarding enforcement mechanisms, referring users to the Branded Content Policy for specifics.

TikTok hosts a Commercial Content Library \cite{tiktok_library_faq_2025}, introduced in July 2023 for DSA Article 39 compliance, through which it aggregates ads and creator-disclosed commercial content with at least one impression. The library contains two sections: the Ad Library (content TikTok is paid to display) and ``Other commercial content'' (content TikTok is not paid to display but which creators have disclosed). The ``Other commercial content'' section depends entirely on creators using the disclosure toggle -- if a creator does not enable it, the content does not appear regardless of its commercial nature \cite{tiktok_library_faq_2025}. 

The Business Help Center \cite{tiktok_content_disclosure_setting_2025} claims TikTok flags videos as commercial content based on financial incentives (URLs, promo codes, hashtags, collaboration mentions, QR codes, overlay ads), brand mentions (brand hashtags, @tags, logos), and product recommendations (calls-to-action such as ``buy now'' or ``shop today''), yet how this operates in practice remains unclear. We identified videos with multi-thousand impressions where creators mentioned brand partnerships in descriptions or in the video itself, yet the platform took no visible action.
\subsection{Manual Content Analysis} 
Second, we identified influencers popular among minors across categories: fitness, fashion, beauty, gaming, and lifestyle. We focused on creators known to have brand deals and examined how they label their content. We manually reviewed 15 influencers and approximately 50 videos.
We found substantial variation both within individual accounts and across creators. Follower size did not determine disclosure practices, meaning larger influencers with hundreds of thousands of followers sometimes used only hashtags (ad, Werbung) or promo codes in descriptions without enabling the toggle. Thus, these videos do not appear in TikTok's Commercial Content Library and lack the grey disclosure label (Paid Partnership) at the bottom of the video. 

We also found creators promoting third-party products who enabled the toggle but selected the wrong category, labelling content as ``Promotional content'' (dedicated to be used by brands or users to indicate their own commercial content) instead of ``Paid partnership'' (dedicated to creators to indicate third party commercial content). It seems that TikTok is not taking any actions to correct these labels over time. Furthermore, we identified variation between TikTok's app and desktop interfaces. Content labelled as "Ad" in-app appeared sometimes as "Sponsored" on desktop and vice-versa.

\section{Algorithmic Audit of TikTok Advertisement System} 

\subsection{Methodology and Experimental Setup}

To empirically assess TikTok’s compliance with Article 28(2) of the DSA, we conducted an algorithmic audit using a sock-puppet approach\footnote{To allow replicability of results presented in this study, the anonymised collected data, altogether with their predicted annotations, are available for research purposes only at Zenodo upon request: \url{https://doi.org/10.5281/zenodo.18879043}; the code used for data analysis is publicly available as a Github repository: \url{https://github.com/kinit-sk/ai-auditology-advertising-and-minor-profiling-tiktok}.}. This methodology allowed us to control for user variables, specifically age and interest, while eliminate undesired confounding factors (such as random variations in gender, location, activity time). As a result, we were able to systematically observe the platform's ad delivery and recommendation behaviour in a naturalistic setting.

\subsubsection{Experimental Design and Agent Configuration} We employed a \textit{paired user study design} to isolate the effect of age on ad delivery. We created paired automated agents (sock-puppets), each consisting of a \textit{minor} agent (aged 16--17) and an \textit{adult} agent (aged 20--21). We avoided the borderline age of 18 (legal majority in all EU Member States except Scotland) to ensure a clear distinction in platform signals. The small age gap was chosen to minimise confounding from generational content preferences, ensuring that differences in recommendations reflect legal status rather than cohort effects.

The gender (\textit{female} or \textit{male}) was assigned to each pair randomly, being the same for minor and adult user (to prevent introducing another undesired confounding factor).

Out of total 4 pairs of agents, each pair was assigned a specific interest profile corresponding to high-engagement topics among adolescents: \textit{beauty}, \textit{fitness}, \textit{gaming} and \textit{politics}. These topics are selected from the main topical groups provided by the TikTok platform in the Explore feature -- out of them, we selected such topics, where the sufficient amount of commercial communication appears.

All agents were configured to access the platform via the \textit{desktop web interface} (Google Chrome). To ensure the application of the DSA's jurisdiction, all traffic was routed through static residential proxies located in \textit{Germany}, establishing a consistent geo-location signal for the platform's compliance mechanisms. Age was signaled to the platform exclusively through the date of birth provided during account registration, without additional behavioural age verification signals. The audit took place during December 2025.

\subsubsection{Interest Seeding Phase} To overcome the ``cold start'' problem inherent in fresh accounts, we implemented a rigorous one-day seeding phase prior to the main data collection phase. Agents began as ``blank slates'' and were trained to acquire their target interest profiles through active engagement.

For each interest profile, we defined a set of domain-specific search queries (e.g., makeup, skincare, or cosmetics for the beauty topic). The agents executed these queries and selected videos from the search results to interact with based on a semantic relevance check. To decide how a user should interact with each of the returned videos, we utilized a \textit{user interaction predictor}. This component employs a Large Language Model (LLM), specifically \textit{GPT-4.1}, to analyze the video metadata (e.g., title and description) and determine whether the topic of the video matches the topic from the user interest profile. The utilized LLM prompt is available in Appendix \ref{appendix:prompt}. This classification model was reused from our previous work \cite{pecher2026algorithmicauditpersonalisationdrift}, where we thoroughly evaluated it. It achieves accuracy of 95-98\% (depending on a specific topic), being sufficient for the auditing purposes.

In user-video topic matches, the agent signalled  ``strong interest'' by watching the video until the end, ``liking'' it, and saving it to bookmarks. This training loop ran until the agent had either watched 25 relevant videos or evaluated 51 candidates, ensuring a consistent baseline of interest history across all agents.

\subsubsection{Data Collection Phase} Following the seeding phase, we conducted the main data collection over a 10-day longitudinal period in December 2025. Each bot interacted with the TikTok for ~1 hour in one session each day, while each adult-minor pair was running in parallel. The duration of ~1 hour corresponds to an average time spent by TikTok users on the platform per day. By running bots contemporaneously, we secured that there are no time-of-day-specific effects that may cause potential undesired biases.

Unlike passive observation studies, our agents maintained their interest profiles through conditional interaction, simulating a user who selectively engages with relevant content while ignoring irrelevant recommendations.

For every video presented in the ``For You'' feed, the system performed a real-time relevance assessment using the same \textit{user interaction predictor} based on the video's textual metadata. If the video matched the agent's assigned interest, the agent executed the engagement routine (watch, like, and bookmark). If the content was unrelated, the agent immediately skipped the video. Crucially, we captured comprehensive metadata---including descriptions, hashtags, and visual frames---for \textit{all} videos appearing in the feed, regardless of whether they were watched or skipped. This ensured we captured the platform's unfiltered recommendation stream and ad delivery logic rather than just the user's interaction history.

\subsubsection{Ad Detection and Classification Pipeline} A core challenge of this audit was identifying not only formal advertisements but also disclosed and undisclosed advertisements. To address this, we developed a multi-modal classification pipeline utilizing a quantized Large Vision-Language Model (LVLM), specifically \textit{Qwen3-VL-4B-Instruct}. The specific prompt used for annotation purposes is available in Appendix \ref{appendix:prompt}. The ad detection and classification pipeline was applied on collected data in the post-audit phase, i.e., the analysis does not need to be done in real-time (in contrast to the user interaction predictor) and thus a deeper content analysis with a larger model was possible. 

For each video, the model analysed three key frames (beginning, middle, and end) to detect the advertisement presence and eventually, classify it into three mutually exclusive ad types based on a strict decision hierarchy:

\begin{enumerate} 
    \item \textit{Formal Ads} -- content where the platform injects a formal overlay label (specifically ``Sponsored'' or ``Ad'') at the bottom of the video frame. 
    \item \textit{Disclosed Ads} -- content where the creator has utilized the platform's disclosure tools, resulting in a ``Paid partnership'' or ``Promotional content'' grey overlay label.
    \item \textit{Undisclosed Ads} -- content lacking any platform-generated disclosure labels but containing strong semantic or visual indicators of commercial intent identified by the model (e.g., verbal product endorsements, visible discount codes, or hashtags). \end{enumerate}

The model also classified ad topic into one of the following options: \textit{beauty}, \textit{fitness}, \textit{gaming}, \textit{politics} or \textit{other}.

This visual-first approach allowed us to detect regulatory compliance failures that text-only analysis would miss, particularly where overlay labels are rendered in the UI but missing from the textual metadata.

\subsubsection{Manual Validation of Ad Detection and Classification Pipeline} To ensure validity, a subset of the automatically annotated videos was manually annotated by the research team to verify the accuracy of the detection logic. For this purpose, to achieve a representative set of data, we applied the stratified sampling strategy on the top of the collected and annotated data. More specifically, for each out of 8 users, we randomly selected 5 videos containing formal, disclosed, undisclosed, and none ads (if there was less than 5 videos in any of these categories, all videos have been selected). 

Utilizing the custom developed tool (see Appendix \ref{appendix:annotation_tool}), two authors of this study independently manually examined the selected videos and provided human-based ad type and topic annotations. They have been subsequently compared with automatic annotations. Specifically for \textit{politics} topic, we observed that both users (a minor as well as an adult) got enclosed during the main audit phase in very specific topical filter bubbles (a significant potion of videos contained AI-generated satire, or videos tackling currently ongoing Pakistani political issues). The classification model did not perform well on this type of content in certain cases, and the videos were also difficult to annotate for human annotators. Therefore, we decided to remove \textit{politics} topic from further analysis.

For the remaining topics, out of 113 manually annotated videos, the automatic annotated ad type was correct in 102 (90.3\%) and 100 (88.5\%) of samples respectively (according to individual annotators). Out of 78/74 videos containing any type of advertisement, the ad topic was correctly annotated in 74 (94.9\%) and 63 (85.1\%). The inter-annotator agreement achieved high rates -- 94.7\% for ad type and 89.2\% for ad topic. A lower agreement for ad topic was caused by videos containing fashion ads, which can be considered as closely relevant and sometimes overlapping with the beauty topic and annotators approached it in some cases differently.

Deeper insight into misclassified samples revealed that the errors occurred approximately equally between each pair of predicted-true labels (see also Figure~\ref{fig:confusion_matrices_annotations}), therefore, no systematic bias has been introduced. Considering the challenging nature of these annotation tasks (especially detection of undisclosed commercial content), we consider the achieved accuracy as sufficient for the purpose of this audit study.

\subsection{Results}
\label{sec:results}

\subsubsection{Metrics and Statistical Analysis Description}
For the purpose of this study, we consider the advertisement to be \textit{personalized} to the user interest profile, if its topic matches a topic present in the used interest profile. Subsequently, a \textit{personalization rate} -- calculated as a proportion of personalized ads out of total ads -- represents a proportion of ads presented to a user that matches their interest profile. However, some number of advertisements may naturally occur in the user feed regardless of their interest profile and such a number can vary accordingly to the popularity and overall prevalence of such advertisements in the platform. To consider this potential bias, we measure explicitly a \textit{profiling effect}. Profiling effect is a percentage point (pp) difference ($\Delta$) between a personalization rate (i.e., a ratio between ads relevant for a user and all ads seen by the user) and a baseline rate (a proportion of ads with the same topic that naturally appeared for users with the same age but different interest profiles). 

To evaluate statistical significance of profiling effect (i.e., whether a difference between personalization and baseline rate is statistically significant), we measure the p-value of proportions z-test.

\subsubsection{Dataset Description}

The outcome of our paired user audit study are data collected by six sock-puppet accounts (differentiated by an age group and an interest) and automatically annotated into four mutually exclusive video categories: formal ads, disclosed ads, undisclosed ads and non-ads and four ad topics: beauty, fitness, gaming, other. Overview of individual accounts, their characteristics and total number of videos and ads are presented in Table \ref{tab:overview_table}. In total, during the 10 days of main data collection phase, the users encountered 7095 videos. Out of them, 1346 (18.97\%) were classified to contain one type of advertisement we are interested in. Adults watched more videos primarily because the content recommended to them was significantly shorter (by ~10 seconds) on average than the content served to minors. Summary of all collected and annotated data are available in Appendix~\ref{appendix:summary_of_data}, Table~\ref{tab:summary_table}.

Summary of all collected and annotated data are available in Appendix~\ref{appendix:summary_of_data}, Table~\ref{tab:summary_table}.

\begin{table}[tbp]
    \caption{Overview of user characteristics and total number of observed videos and ads (formal + disclosed + undisclosed)}
    \label{tab:overview_table}
    \centering
    \resizebox{1\linewidth}{!}{%
    \begin{tabular}{llrrrr}
    \toprule
    \textbf{User} & \textbf{Gender} & \textbf{Age} & \textbf{Total videos} & \textbf{Total ads} & \textbf{Avg video length} \\
    \midrule
    Beauty\textsubscript{Minor} & Female & 16 & 651 & 169 (25.96\%) & 55 seconds \\
    Beauty\textsubscript{Adult} & Female & 21 & 1088 & 246 (22.61\%) & 36 seconds \\
    Fitness\textsubscript{Minor} & Male & 17 & 876 & 45 \enspace (5.14\%) & 36 seconds\\
    Fitness\textsubscript{Adult} & Male & 20 & 2341 & 534 (22.81\%) & 14 seconds \\
    Gaming\textsubscript{Minor} & Female & 16 & 868 & 51 \enspace (5.88\%) & 30 seconds \\
    Gaming\textsubscript{Adult} & Female & 21 & 1271 & 301 (23.68\%) & 30 seconds \\
    \midrule
    Total & & & 7095 & 1346 (18.97\%) \\
    \bottomrule
    \end{tabular}}
\end{table}

\begin{figure}[tbp]
    \centering
    \includegraphics[width=1\linewidth]{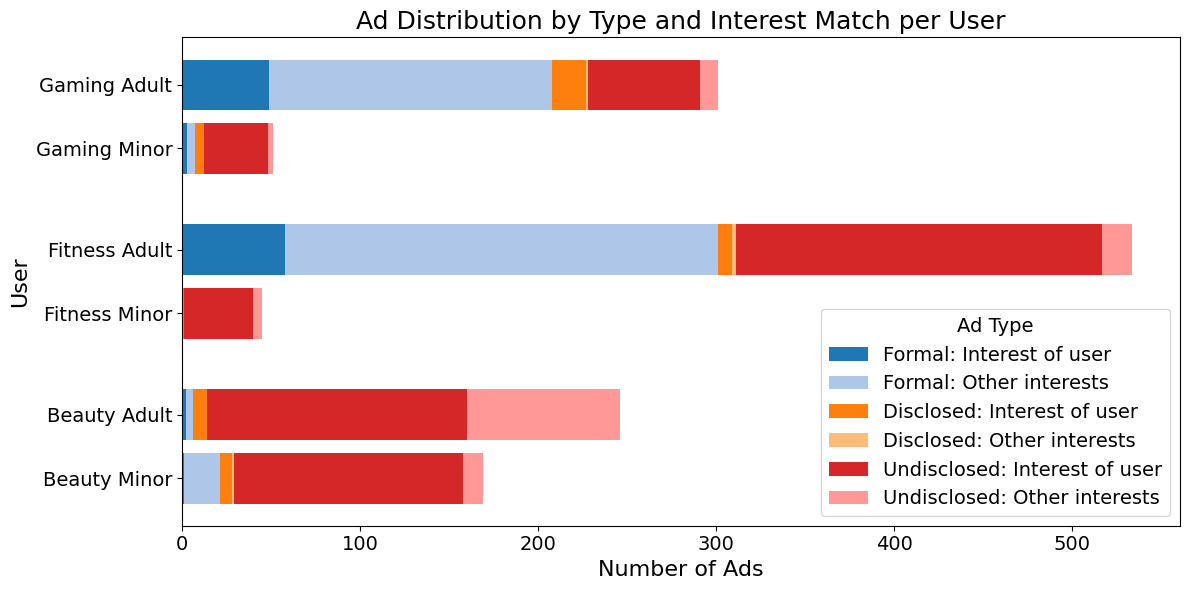}
    \caption{Ad distribution by ad type and ad topic match per user. While minors are recommended less ads in total, such ads are not properly disclosed and are heavily personalized to match their interest.}
    \Description{Ad distribution by ad type and ad topic match per user. While minors are recommended less ads in total, such ads are not properly disclosed and are heavily personalized to match their interest.}
     \label{fig:ad_type_distribution}
\end{figure}

\begin{figure*}[ht]
    \centering
    \subfloat[Formal Ads, Adults\label{ad_topic_distribution_formal_adults}]{
        \includegraphics[width=0.24\textwidth]{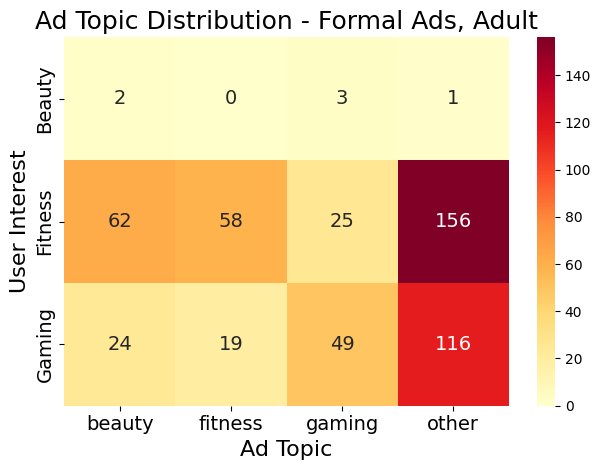}
    }
    \hfill
    \subfloat[Formal Ads, Minor\label{ad_topic_distribution_formal_minor}]{
        \includegraphics[width=0.23\textwidth]{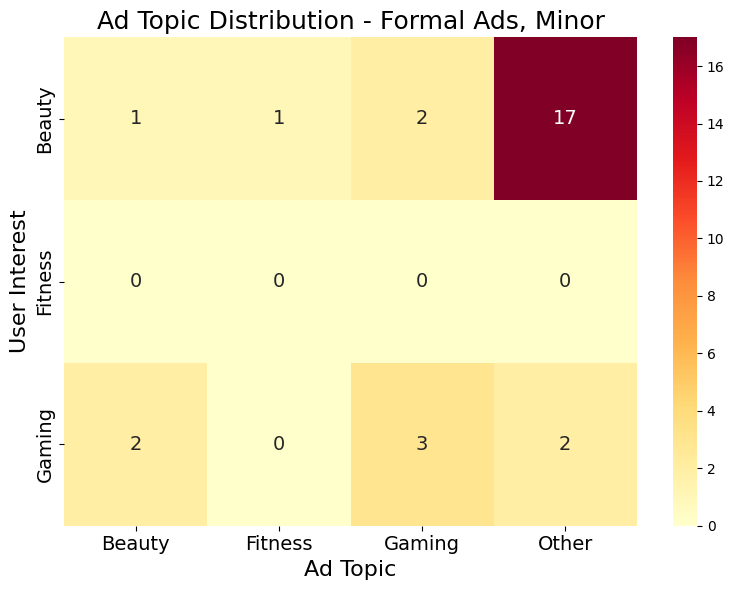}
    }
    \hfill
    \subfloat[Disclosed + Undisclosed Ads, Adults\label{ad_topic_distribution_other_adults}]{
        \includegraphics[width=0.24\textwidth]{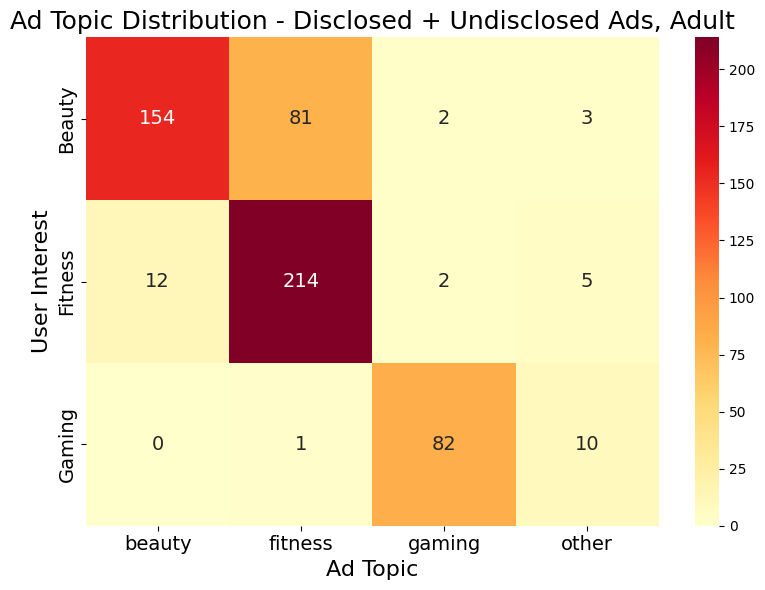}
    }
    \hfill
    \subfloat[Disclosed + Undisclosed Ads, Minor\label{ad_topic_distribution_other_minor}]{
        \includegraphics[width=0.24\textwidth]{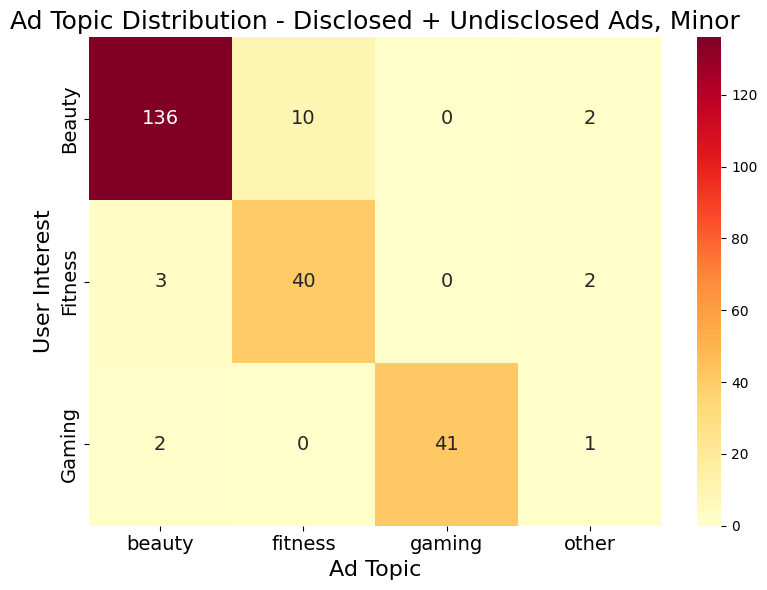}
    }

    \caption{Ad topic distribution. While formal ads distribution indicate no significant personalization, disclosed and undisclosed ads show strong diagonal dominance.}
    \Description{Ad topic distribution.}
    \label{fig:ad_topic_distribution}
\end{figure*}

Such data allowed us to perform a unique investigation across three dimensions: 1) How advertisement types are distributed for adults vs. minors and across different user interest profiles (Figure \ref{fig:ad_type_distribution}); 2) How advertisement topics are distributed for adults vs. minors (Figure \ref{fig:ad_topic_distribution}); and finally 3) An extent to which advertisement are personalized to user interest profiles (Table \ref{tab:formal_ads_personalization} and Table \ref{tab:disclosed_undisclosed_ads_personalization}). As we observed only a limited number of disclosed ads, for the purpose of some analysis we group them together with undisclosed ads (as from the perspective of DSA, they are treated equally -- both fall outside Article 28(2)'s scope and represent creator-driven commercial content differing only in labelling tag).

\subsubsection{Overall Advertising}
As Table \ref{tab:overview_table} clearly shows, advertising exposure (regardless of its specific type) varies substantially by interest category and age group. Minors with fitness and gaming interests are less exposed to advertisements (comprising 4.35\% and 5.18\% of their feeds respectively). Nevertheless, a minor interested in beauty was exposed to a significantly larger amount of ads (23.12\%), even surpassing the paired adult user (20.08\%). This finding can be attributed to a high prevalence of beauty-oriented commercial content on TikTok platform, that is created especially by a considerable number influencers.

A deeper insight into distribution of advertisement types and their match with user profiles (Figure \ref{fig:ad_type_distribution}) further shows that despite minors are exposed to less ads (in absolute as well as relative numbers), such ads comprises particularly undisclosed content that matches their interest.

In the following analysis, we provide a more detailed breakdown of results by each type of advertisement.

\subsubsection{Formal Advertising}
Regarding formal ads, the category explicitly covered by DSA, we observed patterns consistent with the regulation's intent. Adult accounts received substantially more formal advertising than minors (e.g., 301 formal ads for Fitness\textsubscript{Adult} compared to 0 for Fitness\textsubscript{Minor}; 207 for Gaming\textsubscript{Adult}  compared to 7 for Gaming\textsubscript{Minor}). An exception is a beauty topic, where both an adult and a minor received only a relatively low number of formal ads (6 and 21 respectively).

From ad topic distribution (Figure \ref{fig:ad_topic_distribution}, sub-charts (a) and (b)), the proportion of ads matching user interest (i.e., a diagonal dominance) goes up to 21.2\% for adults vs. 14.3\% for minors. Most advertisement presented to minors (and partially also for adults) was classified as ``other'' topic, indicating low personalization of formal ads.

Adult formal ads also exhibited a higher profiling toward user interests as it is in case of minors (Table \ref{tab:formal_ads_personalization}). The profiling effect ranges between 10.39 pp to 16.43 pp (being statistically significant for fitness and gaming topics), e.g., Fitness\textsubscript{Adult} achieved a +10.39pp profiling effect (19.17\% personalization rate compared to 8.88\% baseline rate, p<0.01). Minor accounts, by contrast, showed no consistent interest-based profiling: Beauty\textsubscript{Minor} exhibited a negative effect ($\Delta$ = -23.81 pp), Fitness\textsubscript{Minor} received zero formal ads, and while Gaming\textsubscript{Minor} showed 3 of 8 ads to be interest-matched, the small numbers precludes reliable inference.

These results indicate that TikTok’s formal advertising system operates in compliance with DSA Article 28(2).

\subsubsection{Disclosed Advertising}
Disclosed ads which is understood as content labelled by brands and creators as promotional content or paid partnership, constituted a negligible share of detected advertising across all accounts (both for minors and adults) (Figure \ref{fig:ad_type_distribution}). Among minors, we observed only 8 disclosed ads for beauty, 1 for fitness, and 5 for gaming (Table \ref{tab:summary_table}). This scarcity does not reflect an absence of brand/influencer marketing, it rather reveals a labelling deficit. The disclosure rate (disclosed ads as a proportion of all creator-driven commercial content) was 5.4\% for Beauty\textsubscript{Minor} (8 out of 148), 2.2\% for Fitness\textsubscript{Minor} (1 out 45) and 11.4\% for Gaming\textsubscript{Minor} (5 out of 44). Adult accounts showed similarly low rates. This observation aligns with previous work \cite{auer2025tiktok} providing quantitative evidence that formal disclosure tools for commercial content are rarely used by creators.

Even within this small advertisement type category, personalization was evident: 7 out of 8 disclosed ads for Beauty\textsubscript{Minor} concerned beauty topics and all 5 disclosed ads for Gaming\textsubscript{Minor} were gaming-related.

\subsubsection{Undisclosed Advertising}
Undisclosed advertising is type of ads that are clearly commercial content but lack creator disclosure and the corresponding label. These undisclosed ads constituted the largest advertising category (constituting 55.8\% of all ads) and exhibited personalization intensities far exceeding those observed in formal advertising. The high-prevalence of undisclosed advertisement is in line with the previous work~\cite{doi:10.1287/mksc.2024.0838}, which reported an undisclosed share of commercial communication on Twitter even up to 96\%. Beauty\textsubscript{Minor} received 140 undisclosed ads, of which 129 (92.1\%) matched their interest profile; Fitness\textsubscript{Minor} received 44, with 39 (88.6\%) fitness-related, Gaming\textsubscript{Minor} received 39, with 36 (92.3\%) gaming-related (Table \ref{tab:summary_table}). 

\begin{table}[b]
    \caption{Profiling effect for \textit{formal} ads. \textit{Personalization rate} is a ratio between ads relevant for a user and all ads. \textit{Baseline rate} is determined by proportion of ads with the same topic that naturally appeared for users with the same age but a different interest. \textit{Profiling effect} is a percentage point (pp) difference between these two numbers.}
    \label{tab:formal_ads_personalization}
    \centering
    \resizebox{\linewidth}{!}{%
    \begin{tabular}{lllrr}
    \toprule
    \textbf{User} & \begin{tabular}[x]{@{}r@{}}\textbf{Pers. Rate}\\Pers./Total Ads\end{tabular} & \begin{tabular}[x]{@{}r@{}}\textbf{Baseline Rate}\\Pers./Total Ads\end{tabular} & \begin{tabular}[x]{@{}r@{}}\textbf{Profiling}\\\textbf{Effect}\end{tabular} & \begin{tabular}[x]{@{}r@{}}\textbf{Stat.}\\\textbf{Diff.}\end{tabular}\\
    \midrule
    Beauty\textsubscript{Minor} & 4.76\% (1/21) & 28.57\% (2/7) & -23.81 pp &  \\
    Beauty\textsubscript{Adult} & 33.33\% (2/6) & 16.90\% (86/509) &  16.43 pp &  \\
    Fitness\textsubscript{Minor} & \enspace 0.00\% (0/0) & \enspace 3.57\% (1/28) & -3.57 pp & \\
    Fitness\textsubscript{Adult} & 19.27\% (58/301) & \enspace 8.88\% (19/214) & 10.39 pp & ** \\
    Gaming\textsubscript{Minor} & 42.86\% (3/7) & \enspace 9.52\% (2/21) & 33.34 pp & * \\
    Gaming\textsubscript{Adult} & 23.56\% (49/208) & \enspace 9.12\% (28/307) & 14.44 pp & *** \\
    \bottomrule
    \end{tabular}}
    {\raggedright \footnotesize *** for p < 0.001 (highly significant), ** for p < 0.01 (very significant), * for p < 0.05 (significant) \par}
\end{table}

\begin{table}[b]
    \caption{Profiling effect for \textit{disclosed} and \textit{undisclosed} ads.}
    \label{tab:disclosed_undisclosed_ads_personalization}
    \centering
    \resizebox{\linewidth}{!}{%
    \begin{tabular}{lllrr}
    \toprule
    \textbf{User} & \begin{tabular}[x]{@{}r@{}}\textbf{Pers. Rate}\\Pers./Total Ads\end{tabular} & \begin{tabular}[x]{@{}r@{}}\textbf{Baseline Rate}\\Pers./Total Ads\end{tabular} & \begin{tabular}[x]{@{}r@{}}\textbf{Profiling}\\\textbf{Effect}\end{tabular} & \begin{tabular}[x]{@{}r@{}}\textbf{Stat.}\\\textbf{Diff.}\end{tabular}\\
    \midrule
    Beauty\textsubscript{Minor} & 91.89\% (136/148) & \enspace 5.62\% (5/89) & 86.27 pp & *** \\
    Beauty\textsubscript{Adult} & 64.17\% (154/240) & \enspace 3.68\% (12/326) &  60.49 pp & *** \\
    Fitness\textsubscript{Minor} & 88.89\% (40/45) & \enspace 5.21\% (10/192) & 83.68 pp & *** \\
    Fitness\textsubscript{Adult} & 91.85\% (214/233) & 24.62\% (82/333) & 67.23 pp & *** \\
    Gaming\textsubscript{Minor} & 93.18\% (41/44) & \enspace 0.00\% (0/193) & 93.18 pp & *** \\
    Gaming\textsubscript{Adult} & 88.17\% (82/93) & \enspace 0.85\% (4/473) & 87.32 pp & *** \\
    \bottomrule
    \end{tabular}}
    {\raggedright \footnotesize *** for p < 0.001 (highly significant), ** for p < 0.01 (very significant), * for p < 0.05 (significant) \par}
\end{table}

For ad topic distribution, we report diagonal dominance for disclosed and undisclosed ads combined (Figure \ref{fig:ad_topic_distribution}, sub-charts (c) and (d)). For minors, diagonal dominance was 91.6\% (217 of 237 ads) -- to compare, a diagonal dominance for formal ads was only 14.3\%. Figure \ref{fig:ad_topic_distribution} visualizes this contrast by highlighting that disclosed and undisclosed ads cluster tightly on the diagonal for both minor and adults, whereas minor formal ads disperse across the ``other'' topical category.

Analysing undisclosed ads separately yields 91.5\% (204 of 223); within disclosed ads alone, interest-matching was similarly concentrated (e.g., 7 of 8 for Beauty\textsubscript{Minor}, though the small sample size limits interpretation). This confirms that combining disclosed and undisclosed ads does not obscure category-specific patterns.

For disclosed and undisclosed ads, profiling effects for minors ranged from +83.68 pp to +93.18 pp (See Table \ref{tab:disclosed_undisclosed_ads_personalization}). Beauty\textsubscript{Minor} showed 91.89\% personalization rate versus a 5.62\% baseline rate ($\Delta$ = +86.27 pp, p<0.001); Fitness\textsubscript{Minor} showed 88.99\% versus 5.21\% baseline ($\Delta$ = +83.68 pp, p<0.001) and Gaming\textsubscript{Minor} showed 93.18\% versus 0.00\% ($\Delta$ = + 93.18 pp, p<0.001). 

The results show that minors received stronger interest-based targeting through disclosed and undisclosed ads (+83.68 pp to +93.18 pp) than adults received through formal ads (+10.39 pp to +16.43 pp, Table \ref{tab:formal_ads_personalization}). These profiling effects are 5-8 times larger than adult formal advertising profiling effects that DSA Article 28(2) permits.

\section{Discussion and Policy Recommendations}

Our results show two central findings. First, the majority of advertising falls into the undisclosed category, suggesting that creators and brands are not using the available labelling tools. However, even in this case, TikTok officially complies with Article 26(2) of DSA because it in fact provides the functionality for creators to declare commercial content, but the platform's obligation ends there. When creators fail to label such content through the provided tools, the platform bears no direct obligation to intervene -- neither detects the omission nor corrects it. Second, it is precisely these undisclosed ads that exhibit the strongest profiling toward user interests. TikTok demonstrates formal compliance with Article 28(2) by shielding minors from profiled formal ads, but minors still remain exposed to highly profiling-based advertising that the regulation does not recognize as such. 

The pattern across categories reinforces this finding. Formal ads do not show evidence of profiling toward minors, which is consistent with the requirements of the DSA. We observed a slight increase in profiling toward minors with disclosed ads, where creators and brands used corresponding labelling tools. The vast majority falls into the undisclosed category. Minors with specific interest profiles received a significantly higher number of undisclosed ads aligned with those interests at higher rates (ranging from +83.68 to +93.18 percentage points) than other minors with different interests. The finding related to the beauty topic deserves particular attention beyond its statistical significance. The literature consistently links targeted exposure to appearance-related commercial content with negative body image and diminished self-esteem among adolescent girls \cite{ChoukasBradley2022PerfectStorm, Blackburn2024ForYou}. Therefore, a 16-year-old girl algorithmically fed undisclosed beauty advertisements at a 92.1 percent personalization rate represents a documented harm pathway.

The gap appears to be structural. Article 3(r)'s narrow definition of advertisement excludes influencer marketing and brand self-promotion from Article 28(2)'s scope. This framing is notably narrower than  Directive 2006/114/EC, which defines advertising as ``the making of a representation in any form in connection with a trade, business, craft or profession in order to promote the supply of goods or services''. This definition is based on commercial purpose rather than payment structure \cite{EU_Directive_2006_114_EC}. It is also narrower than the E-Commerce Directive's (Directive 2000/31/EC) definition of commercial communication, which covers any form of communication for promotion with aim of commercial gain. TikTok thus demonstrates formal compliance while the vast majority of advertising and its subsequent profiling reaches minors and operates outside the prohibition's reach.

Based on our findings, we propose the following four recommendations. 

\begin{enumerate}
    \item First, we advocate for expansion of the definition of advertisement in EU law to encompass advertising by commercial purpose rather than payment structure, which would as a result include influencer and brand promotional content in line with the spirit of definitions of advertising and commercial communications across EU legislation.
    \item Second, labelling obligations should be strengthened so that platforms bear direct responsibility for detecting and correcting missing labels on commercial content, not merely providing functionality for creators and brands to self-label. Our classification model achieved accuracy as high as 88.5/90.3\% using only publicly observable features, suggesting that platforms' internal models should perform even better. Although VLOPs should detect these issues through risk-assessment obligations, fulfilment relies on thorough auditing and enforcement -- and this issue has not yet been tackled \cite{EBDS2025FirstReport}. Alongside platform responsibility, creators should be subject to a graduated accountability mechanism: where a platform detects unlabelled commercial content, it should flag it, apply the appropriate label, and warn the creator, with repeated failures triggering content removal.
    \item Third, any legislative instrument expanding the definition of advertisement to encompass influencer and brand promotional content must include a corresponding prohibition on profiling-based targeting of minors, ensuring that the protections currently limited to formal advertisements extend to all commercial content regardless of its channel of delivery.
    \item Finally, compliance with these obligations must be independently verified. In our prior work  \cite{SolarovaEtAl2026DSA}, we demonstrated that conventional additional audit methodologies are inadequate for assessing dynamic algorithmic systems. Annual, independent DSA audits should incorporate algorithmic audit mechanism, such as the methodology employed here, to verify whether ads, under this expanded definition, reach minors through profiling. 
\end{enumerate}

There are positive aspect to minors being in the digital environment but they should be able to be present there without being exploited, manipulated or exposed to harmful effects against which they cannot defend themselves the same way adults do. Our findings demonstrate that current definitional boundaries render this vision largely symbolic. The legal precedent exists, the enforcement methodology exists, too. What remains is the regulatory and policy decision to align them.

This study has some inevitable limitations. Conducting the sock-puppet algorithmic auditing studies is a challenging task. Deploying sock-puppet agents in a real-world social media environment is technologically challenging (e.g., to prevent a platform from banning sock-puppet accounts, to realistically simulate user's location, browser or interactions) as well as demanding on costs (e.g., running GPU-powered machines capable analysing videos in real time) and on human oversight (during the audit, the agents were under continuous monitoring). This limits the possible scope of such studies (in our case, it was 4 topics, 1 location, 10 days of the main data collection phase). Furthermore, the simulated user profiles expressed a single interest topic, whereas real adolescents typically exhibit multiple overlapping interests, meaning the observed profiling effects may not fully capture the complexity of real-world algorithmic behaviour. Future studies should also consider using longer-established accounts, as platforms calibrate recommendations over time and fresh accounts may not fully reflect how the system operates for  users with a longer interaction history.

The utilized automatic labelling is also naturally imperfect and assigning ad type/topic labels for some samples may be subjective. However, the accuracy and inter-annotator agreement showed that the data quality is high enough taking the purpose of the study into consideration. Also, the observed differences between minor and adult behaviour are so significant, that they cannot be explained by potential false positives introduced by automatic labelling. Therefore, we consider the collected and annotated data to be sufficiently representative and the findings derived from them to be generalizable also to additional topics or locations.

\section{Conclusions}

Minors are spending increasing amounts of time on social media, which raises well-documented concerns about their development and mental health. Meanwhile, social media platforms have evolved into environments which are heavily shaped by commercial content. To protect minors, who are especially vulnerable to persuasive, personalized marketing communication, the EU's legislation -- Digital Service Act (DSA) -- bans profiling of minors for advertising purposes. However, verifying compliance is challenging and requires behavioural approach. Despite a growing body of algorithmic audits, compliance of social media platforms with minor-protection obligations in ad delivery systems remained unexamined.

To fill this gap, we conducted an algorithmic audit of all major types of advertising types on TikTok to examine whether minors are subject to profiling-based delivery of ads despite the DSA's ban. Through algorithmic auditing we deployed sock-puppet agents simulating adult-minor pairs, with matching interest profiles. Collected content was automatically classified into four video categories: formal ads, disclosed ads (labelled paid partnership or promotional content), undisclosed ads (unlabelled paid partnership/promotional content) and non-ads (classic content); as well as their topics.

The results are concerning. On one hand, formal ads are shown less frequently to minors than to adults and show no evidence of profiling toward minors. This indicates TikTok's formal compliance with DSA Article 28(2). On the other hand, disclosed and undisclosed ads, which comprise a majority of commercial content delivered to minors, show very strong personalization rates and profiling effects. These findings reveal a crucial blind spot: the DSA's narrow definition of advertisement, grounded in payment structure rather than commercial purpose, leaves the dominant forms of commercial exposure minors face practically unregulated. To this end, we provide four policy recommendations to close this gap, and call for their urgent consideration given the scale of commercial exposure minors experience without adequate protection. 

In our future work, we plan to cover additional DSA obligations, replicate the audit over time, and expand to additional countries and platforms.

\section*{Ethical Considerations Statement}
The research presented in this paper was done as a part of the research project, which obtained approval from the organisational Ethics Committee (decision as of December 17, 2024). To minimise any potential legal and ethical issues, we directly involve legal and ethics experts as part of this project. Researchers and research engineers conducting this auditing study also participated in four ethics assessment workshops together with ethics and legal experts, where relevant ethical and legal challenges have been identified and appropriate mitigations proposed.

The execution of sockpuppeting audits requires creating automated bots and using them for data collection, which is a potential violation of the terms of service of the social media platforms. However, this breach of ToS is permitted by Article 40 (12) of the EU Act on Digital Services (DSA) if the research concerns systemic risks. This work directly addresses such a systemic risk by the assessment of social media platforms compliance with obligations imposed by legislation, specifically prohibiting profiling-based advertising to minors stated by the Article 28(2) of DSA, as foreseen by Recital 83 of the DSA. Second, the interaction of the bots with the content on the platform may impact the platform and society (e.g., increasing the view or like count). However, we minimise the number of bots that we run. When it comes to data, we collect only publicly available metadata. 

To mitigate potential biases and inaccuracies inherent in the Large Vision-Language Model (LVLM) used for advertisement classification, we implemented a multi-layered validation process. This included both ad-hoc and systematic manual audits of dataset subsets. Data failing to meet accuracy benchmarks were excluded, and we have reported the estimated error rates accordingly. To prioritize ethical standards and researcher well-being, all manual annotations were conducted solely by the study’s authors, following expert ethical guidelines.

\section*{Generative AI Usage Statement}

Generative AI has not been used to generate any text for this publication. LLMs were used solely for the purpose of grammar and stylistic improvements.

\begin{acks}
This work was partially funded by the EU NextGenerationEU through the Recovery and Resilience Plan for Slovakia under the project \textit{AI-Auditology}, No. 09I03-03-V03-00020.
\end{acks}

\bibliographystyle{ACM-Reference-Format}
\bibliography{bibliography}

\appendix

\section{Prompts Utilized for Automatic Data Annotation}
\label{appendix:prompt}

\lstset{
  basicstyle=\ttfamily\footnotesize,
  breaklines=true,
  frame=single,
  columns=fullflexible
}

The prompt used by LLM (\textit{GPT-4.1}) as a part of the \textit{user interaction predictor}:

\begin{lstlisting}
Your task is to determine the topic and the stance of the video given its available metadata. The only possible answers for the topic is: 1) beauty, 2) fitness, 3) gaming, 4) politics. For stance, the possibilities are: 1) support; 2) oppose. Use only these possibilities when answering.

Here is a closer description for the topics and their stances:

Beauty should be given to anything that is related to beauty or personal care and related products.
Fitness should be given to anything that is related to fitness, exercise, workouts and related products.
Gaming should be given to anything that is related to video games or computer hardware.
Politics should be given to anything that is related to politics, national policies, elections and world events. For example politics related to economy, employment, war and peace, housing, education.
Other should be given to anything that is not related to the previous topics.

In this case, the stance does not matter, so always put support.

Provide the answer in structured form that looks like this:

Topic: beauty/fitness/gaming/politics/other
Stance: support/oppose

Here is the video information.
Author: {author_name}
Video description and hashtags: {video_description_and_hashtags}
Video transcript: {video_transcript}
\end{lstlisting}

\newpage

The prompt used by LVLM (\textit{Qwen3-VL-4B-Instruct}) as a part of the \textit{ad detection and classification pipeline}:

\begin{lstlisting}
You are an assistant that identifies commercial content, both disclosed and undisclosed, in TikTok videos for regulatory compliance audits.
Your task is to analyze a screenshot from a short-video platform and classify it into specific categories.

INSTRUCTIONS:
1.  Detection: Determine if the image contains an advertisement.
2.  Type Classification: If it is an ad, classify the TYPE based on visual indicators:
    - Formal: Look for platform-injected labels "Sponsored" or "Ad" displayed on the bottom of the video.
    - Influencer: Look for creator-injected labels "Paid partnership" or "Promotional content" displayed on the bottom of the video,
    - Other: Look for videos with commercial content that lacks proper disclosure. Look for: product names or prices or discount codes mentioned/shown, brand endorsements by the creator, or promotional hashtags (e.g., #ad, #partnership, #collaboration) used without corresponding platform disclosure labels.
3.  Topic Classification: Classify the primary content of the ad into exactly one of these categories:
    - `beauty` (makeup, skincare, cosmetics, clearskin, kbeauty, glasskin)
    - `fitness` (abs, workout, gym, sports, health, nutrition, gymtok, supplements)
    - `gaming` (video games, consoles, streamers, gamer, gaming, gamerlife)
    - `politics` (political news, political debate, voting, politicians, breaking news)
    - `other` (anything else)

RESPONSE FORMAT:
Reply with valid JSON only.
IMPORTANT: Do NOT use double quotes " inside your string values. Use single quotes ' instead.

Example: "reasoning": "The text says 'Buy now' which implies..."

{
  "is_ad": boolean,
  "ad_type": "formal" | "influencer" | "other" | null,
  "ad_topic": "beauty" | "fitness" | "gaming" | "politics" | "other" | null,
  "visual_indicators": ["list", "of", "labels", "found"],
  "reasoning": "brief explanation"
}
\end{lstlisting}

\section{Summary of collected data}
\label{appendix:summary_of_data}

Table \ref{tab:summary_table} provides a summary of all data collected during the main audit phase. These numbers provide necessary basis for all calculations, tables and figures presented in the Section \ref{sec:results}.

\begin{table}[htbp]
\caption{Summary table of data collected during the audit.}
\label{tab:summary_table}
\resizebox{1\linewidth}{!}{
\begin{tabular}{lrrrrrr}
\toprule
Interest & beauty & beauty & fitness & fitness & gaming & gaming \\
Age Group & minor & adult & minor & adult & minor & adult \\
Gender & female & female & male & male & female & female \\
\midrule
Total Records & 651 & 1088 & 876 & 2341 & 868 & 1271 \\
Non Ad Records & 482 & 842 & 831 & 1807 & 817 & 970 \\
Ads Detected & 169 & 246 & 45 & 534 & 51 & 301 \\
Avg Video Length & 54.8 & 35.6 & 35.8 & 14 & 30 & 30 \\
\midrule
Formal Ads & 21 & 6 & 0 & 301 & 7 & 208 \\
Formal Ads / Beauty & 1 & 2 & 0 & 62 & 2 & 24 \\
Formal Ads / Fitness & 1 & 0 & 0 & 58 & 0 & 19 \\
Formal Ads / Gaming & 2 & 3 & 0 & 25 & 3 & 49 \\
Formal Ads / Other & 17 & 1 & 0 & 156 & 2 & 116 \\
\midrule
Disclosed Ads & 8 & 8 & 1 & 10 & 5 & 20 \\
Disclosed Ads / Beauty & 7 & 8 & 0 & 1 & 0 & 0 \\
Disclosed Ads / Fitness & 0 & 0 & 1 & 8 & 0 & 0 \\
Disclosed Ads / Gaming & 0 & 0 & 0 & 0 & 5 & 19 \\
Disclosed Ads / Other & 1 & 0 & 0 & 1 & 0 & 1 \\
\midrule
Undisclosed Ads & 140 & 232 & 44 & 223 & 39 & 73 \\
Undisclosed Ads / Beauty & 129 & 146 & 3 & 11 & 2 & 0 \\
Undisclosed Ads / Fitness & 10 & 81 & 39 & 206 & 0 & 1 \\
Undisclosed Ads / Gaming & 0 & 2 & 0 & 2 & 36 & 63 \\
Undisclosed Ads / Other & 1 & 3 & 2 & 4 & 1 & 9 \\
\bottomrule
\end{tabular}}
\end{table}

\begin{figure}[htbp]
    \centering
    \subfloat[Ad Type - Annotator 1]{\includegraphics[width=0.49\linewidth]{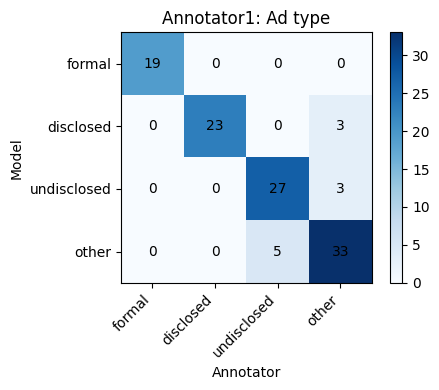}\label{fig:type_a1}}
    \hfill
    \subfloat[Ad Type - Annotator 2]{\includegraphics[width=0.49\linewidth]{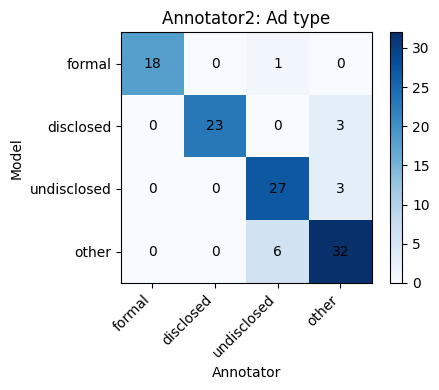}\label{fig:type_a2}}

    \subfloat[Ad Topic - Annotator 1]{\includegraphics[width=0.49\linewidth]{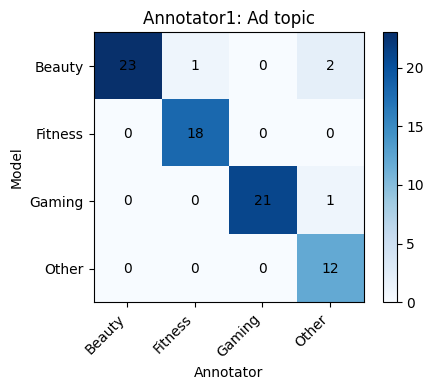}\label{fig:topic_a1}}
    \hfill
    \subfloat[Ad Topic - Annotator 2]{\includegraphics[width=0.49\linewidth]{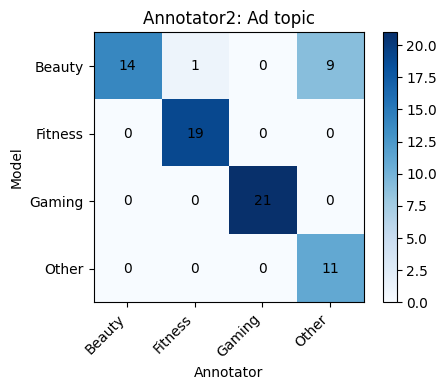}\label{fig:topic_a2}}

    \caption{Predicted-true confusion matrices for ad type and ad topic labels for each of human annotators.}
    \label{fig:confusion_matrices_annotations}
\end{figure}

\section{Tool for Examination and Manual Annotation of Collected Data}
\label{appendix:annotation_tool}

\begin{figure*}
    \centering
    \includegraphics[width=1\linewidth]{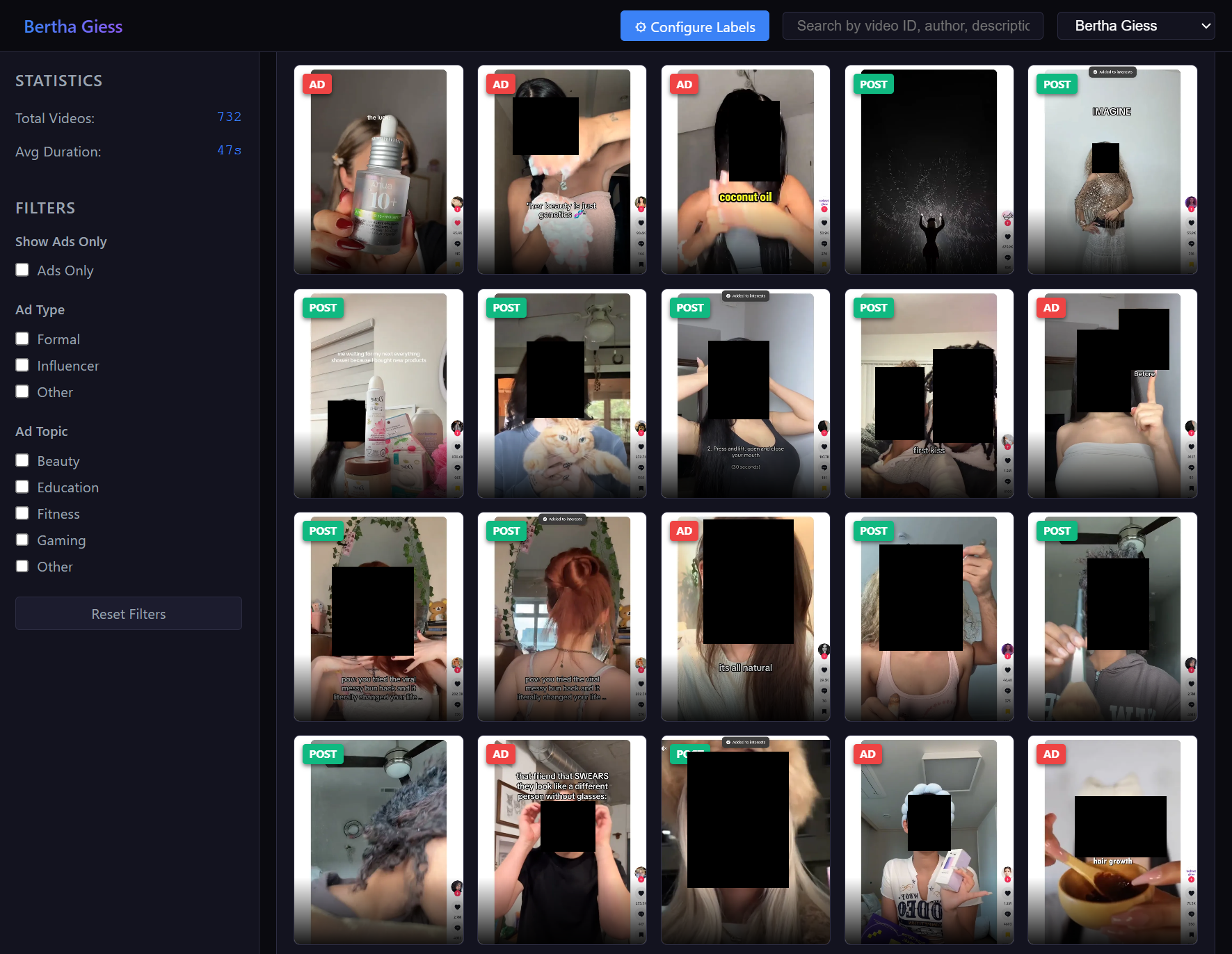}
    \caption{Screenshot of the custom-developed tool providing an overview of the collected data. The displayed videos were presented to a user Beauty\textsubscript{Minor} (a 16-year-old minor with an expressed interest in beauty). The tags determine whether the annotation model determined any kind of advertisement or not. The faces and personal identifying information were blurred or anonymised to preserve creators' privacy.}
    \Description{Screenshot of the custom-developed tool providing an overview of the collected data. The displayed videos were presented to a user Beauty\textsubscript{Minor} (a 16-year-old minor with an expressed interest in beauty). The tags determine whether the annotation model determined any kind of advertisement or not. The faces and personal identifying information were hidden or anonymised to preserve creators' privacy.}
    \label{fig:annotation-tool_overview}
\end{figure*}

\begin{figure*}
    \centering
    \includegraphics[width=1\linewidth]{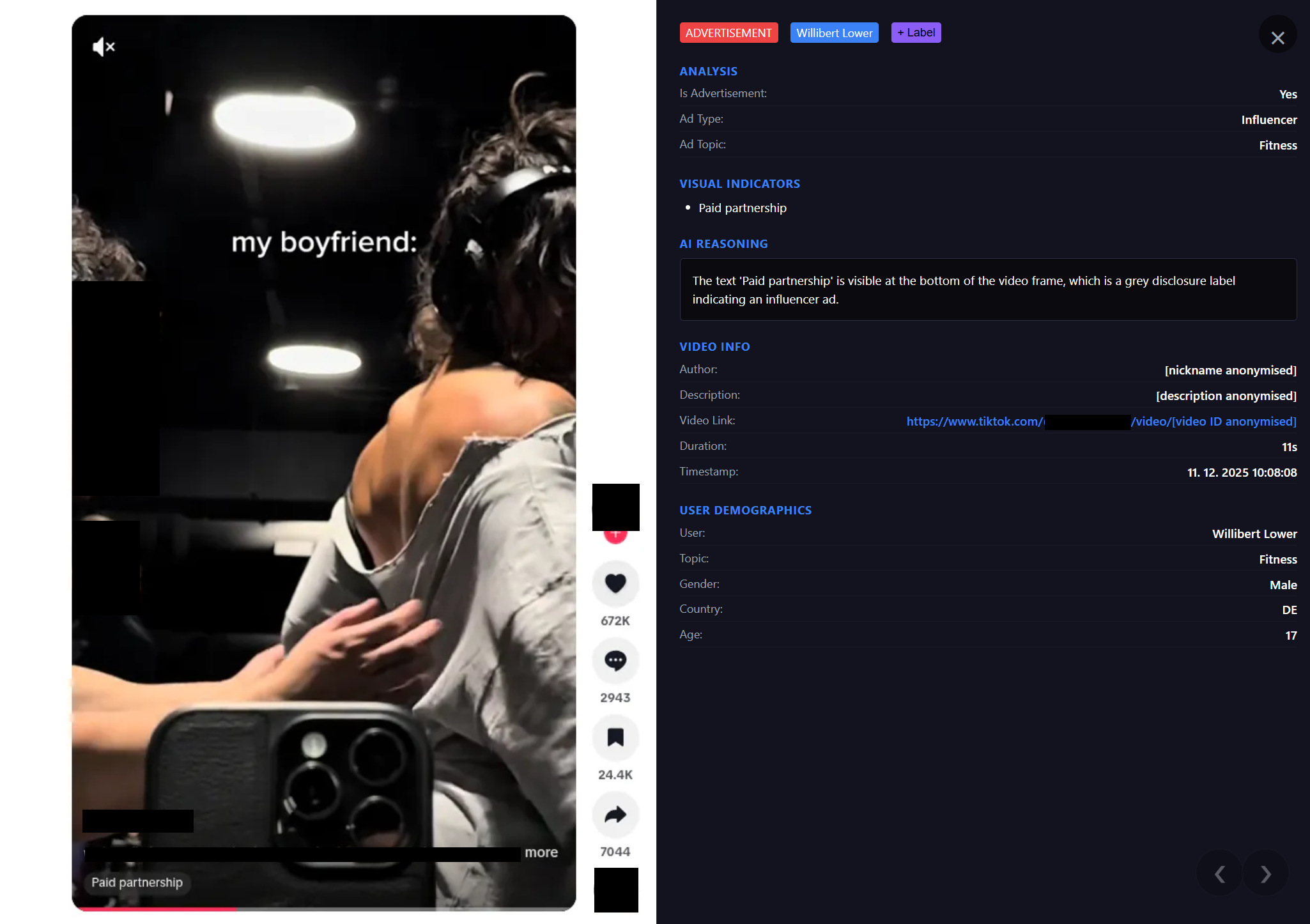}
    \caption{Screenshot of the custom-developed tool providing a detail of the video observed by a user. The displayed video was presented to a user Fitness\textsubscript{Minor} (a 17-year-old minor with an expressed interest in fitness). The tool displays automatic content analysis (including the reasoning of the model) together with the screenshot and URL to the video, streamlining the manual annotation process. The displayed video represents an example of paid partnership content. The faces and personal identifying information were hidden or anonymised to preserve creators' privacy.}
    \Description{Screenshot of the custom-developed tool providing a detail of the video observed by a user. The displayed video was presented to a user Fitness\textsubscript{Minor} (a 17-year-old minor with an expressed interest in fitness). The tool displays automatic content analysis (including the reasoning of the model) together with the screenshot and URL to the video, streamlining the manual annotation process. The displayed video represents an example of paid partnership content. The faces and personal identifying information were blurred or anonymised to preserve creators' privacy.}
    \label{fig:annotation-tool_detail}
\end{figure*}

For the purposes of initial examination and manual annotation of data collected during the algorithmic auditing study, we have developed a custom tool visualizing the collected data. After selecting the corresponding simulated user, it allows to display an overview of TikTok videos displayed to a user, with options to filter the content according to the ad type and topic (Figure \ref{fig:annotation-tool_overview}). By selecting a specific video, it provides a video detail page showing the selected frame from the video, automatic content analysis as well as additional content metadata collected during the study (Figure \ref{fig:annotation-tool_detail}).

The tool served during this study two purposes: 1) it allows us to effectively examine the collected data and obtain the first insights, as well as 2) it provides us the mean to manually evaluate the accuracy of automatic data analysis.

\end{document}